\documentstyle[12pt]{article}
\begin{document}
\baselineskip=24pt

\title{Violent Relaxation of Spherical Stellar Systems}

\author{Motokazu {\sc Takizawa} and Shogo {\sc Inagaki} \\ {\it Department of Astronomy, Faculty of Science, Kyoto University,} \\ {\it Sakyo-ku, Kyoto 606-01}}
\date{(Received \ \ \ \ \ \ \ \ ;accepted \ \ \ \ \ \ \ \ )}

\maketitle

\newpage
\begin{center}
{\bf Abstract}
\end{center}
Violent relaxation process of  spherical stellar systems 
is examined by numerical simulations of shell model. The collapse of 
uniform density sphere both with and without external force is investigated.
It is found that time variation of mean gravitational potential field induced
by external force makes the system closer to Lynden-Bell distribution.
Our result imply that systems which are gravitationary affected by
other systems may approach the Lynden-Bell distribution.


\newpage
\section{Introduction}

Violent relaxation is a mechanism proposed to explain a similarity and
smoothness of the observed light distribution of elliptical galaxies. This
relaxation process is caused by explicitly time-dependant mean
gravitational potential field. Its typical timescale is much shorter than
two-body relaxation time and comparable with dynamical time.

According to Lynden-Bell (1967), it is expected that the coarse-grained
entropy of systems becomes maximum through violent relaxation in ideal cases
and that Lynden-Bell distribution can be established. The Lynden-Bell distribution
is given by
 \begin{eqnarray}
  \bar{f} ( \varepsilon ) = \eta \frac{ \exp \{ - \beta ( \varepsilon - \mu ) \} }{ 1 + \exp \{ - \beta ( \varepsilon - \mu ) \} },
 \end{eqnarray}
where $\bar{f} ( \varepsilon )$ is coarse-grained distribution function,
$\varepsilon$ is energy per unit mass, and $\eta, \beta, \mu$ are
constants. In almost all cases, degeneracy can be neglected  $( \bar{f}
( \varepsilon ) \ll \eta )$, so Lynden-Bell distribution is
approximated by the Maxwell-Boltzmann distribution:
 \begin{eqnarray}
 \bar{f} ( \varepsilon ) = A \exp ( - \beta \varepsilon ),
 \end{eqnarray}
where $ A = \eta \exp ( \beta \mu ) = {\rm constant.}$
However, in numerical simulations so far of one dimensional
gravitational N-body systems (sheet systems), they did not reach
Lynden-bell distribution through violent relaxation ( Cuperman et al. 1969;
Hohl and Campbell 1968; Goldstein et al. 1969; Yamashiro et al. 1992).
It remains unclear why Lynden-Bell distribution cannot be
obtained in these simulations. Moreover it has not been fully investigated what
distributions are obtained in other gravitational N-body systems
(shell systems, particle systems) through violent relaxation.

It is widely believed that the Lynden-Bell distribution is not realized in
N-body simulations (e.g., Funato et al. 1992a,1992b). One of the reasons
is that the fluctuations in the gravitational potential decays before
the relaxation process is completed. To investigate if this is true, we add
artificial time (and space) varying potentials and see if the Lynden-Bell
distribution is established. As a first step, we adopt simple spherical
symmetric models since spherical symmetric models are more realistic than 
sheet models. More realistic models will be studied in subsequent papers.

\section{Numerical Method}

In our simulations, shell models (H\'enon 1964) are adapted. In spherical
symmetric stellar systems, distribution function can be expressed as
$f(r,u,v,t)$, where $r$ the radial distance, $u$ the radial
velocity, $v$ the tangential velocity, and $t$ is the time, respectively. 
So distribution function is represented by $N$ points in $(r,u,v)$ space and
each point evolves according to the equations of motion.
 \begin{eqnarray}
  \frac{dr_i}{dt} &=& u_i, \\
  \frac{du_i}{dt} &=& \frac{A_i^2}{r_i^3} - \frac{G M_i}{r_i^2} - \frac{\partial \phi_{ex} (r, t)}{\partial r},
 \end{eqnarray}
where $G$ the gravitational constant, $A_i = r_i v_i = {\rm constant}$
(with respect to time) is the angular momentum of each shell, $M_i$ the mass interior
to $r_i$, and $\phi_{ex} (r,t)$ is the potential of external force.

We set $G = M$(total mass)$ = 1$ and $N=10000$ . And each shell has the
same mass $( = 1/N )$ . The integration is carried out 
using a leap-frog method.

When we set $\phi_{ex} (r,t) = 0$, we set integration time-step small enough
for the maximum error of total energy to be of the order
of $0.01 \%$ after $100 t_{cr}$ . $t_{cr}$ is  the crossing time which is 
typical timescale for shells to cross typical radius of the system. In our
simulations, $t_{cr} \sim 1$ in this system of units. When 
$\phi_{ex} (r,t) \neq 0$, we set integration time-step same as the model whose
initial condition is common.

\section{Models}

 We calculate collapses of uniform density sphere whose initial virial ratio
is 0.5 and 0.1. In some models  external gravitational potential  field 
are added after systems are settled down to study equilibrium states. 

Initial densities and velocity distributions are common in all models.
 \begin{itemize}
 \item $\rho (r) = {\rm const.}$  $(0 \le r \le 2)$ .
 \item Both $u$ and $v$ obey isotropic Maxwell distribution.
 \end{itemize}

 For simplicity, we assume that both time and spatial dependence of external 
potential field is sinusoidal. We set $\phi_{ex}$ as follows,
 \begin{equation}
 \phi_{ex} (r, t) =  \left\{
                       \begin{array}{@{\,}ll}
                        A \sin ( 2 \pi t / P ) \sin ( 2 \pi r / \lambda ) & \mbox{($20 \le t \le 20 + 5 P$)} \\
                        0 & \mbox{(other)}
                       \end{array}
	                     \right. \label{eq:phiex}
 \end{equation}

 In (\ref{eq:phiex}) $P$ and $\lambda$ are parameters.
 We set external potential to survive for much shorter time 
than two-body relaxation time which is comparable to $N t_{cr}$ ( see appendix 1 ).
We set typical intensity of external potential comparable to that of
inertial potential. In our simulations, we set A=0.5 when initial 
virial ratio is 0.5, and A=1.36 when initial virial ratio 0.1.

 In table 1 the parameters for each models are listed. In this table both $P$
and $\lambda$ are normalized in the units of $\bar{R}$ and $\bar{t}$,
which are typical 
lengthscale and time scale of the post collapsed system. In models whose 
initial virial ratio is 0.5, $\bar{R} = 1.34$ and $\bar{t} = 1.53$, and 
in models whose initial virial ratio is 0.1, $\bar{R} = 0.494$ and $\bar{t} =0.345$ 
( see appendix 2 ). Models are summarized in table 1.

\section{Results}

In models A-0 and B-0, the system approached stationary states within 
a few crossing times and variation of virial ratio and variation of 
distribution function was damped(figure 1 and figure 2). In figure 2,
the mean value of the virial ratio at the stationary stages is 
greater than $1.0$ (about $1.35$) because the existence of 
shells which have positive energies cannot be negligible.

\begin{tabular}{ccc}
 \cline{1-1} \cline{3-3}
 Figure 1 & \ \ \ \ \ \ \ & Figure 2 \\ 
 \cline{1-1} \cline{3-3}
\end{tabular}

In other models, the systems also approached stationary states within a few 
crossing time after variation of external potential stopped. 
In table 2 ratio of mass whose energy is more than zero in stationary states 
is listed in each model.

In figures 3, figures 4 and figures 5, distribution function of stationary 
states of model A-0, model B-0 and model A-1 are shown by the solid lines.
We derive distribution function from the simulation data as follows. 
We divide energy space between $\varepsilon = \varepsilon_{min}$ and
$\varepsilon = 0$ into $100$ bins, where $\varepsilon_{min}$ is the minimum 
value of the potential energy of the system. 
Therefor the $i$-th bin corresponds
to the energy range between $\varepsilon_i -  \Delta \varepsilon/ 2  $ and
$\varepsilon_i + \Delta \varepsilon / 2  $,
where $\varepsilon_i = \varepsilon_{min} + (i-1/2) \Delta \varepsilon$ 
and $\Delta \varepsilon =  - \varepsilon_{min}/100$.
Then we count the number of shells contained the $i$-th bin and it is divided
by the phase space volume which corresponds to the energy range to get the
$\bar{f} (\varepsilon_i)$. Anisotropy of velocity distribution cannot harm
this procedure because energy of each shell doesn't depend on the direction 
of velocity but only on the absolute value.

\begin{tabular}{ccccc}
 \cline{1-1} \cline{3-3} \cline{5-5}
 Figure 3 & \ \ \ \ \ \ \ & Figure 4 & \ \ \ \ \ \ \ & Figure 5 \\
 \cline{1-1} \cline{3-3} \cline{5-5}
\end{tabular}

We estimate the deviation from Lynden-Bell distribution of each model 
as follows.
Unfortunately, spherical stellar systems with Maxwell distribution 
(isothermal distribution) have infinite mass. So the distribution expected
after violent relaxation in real spherical stellar systems is not Maxwellian
itself, but that of the King model which has 'cut-off' at higher 
energy end. So distribution functions in stationary states are fitted 
by the form,
 \begin{eqnarray}
 \bar{f}( \varepsilon ) = \rho_1 ( 2 \pi \sigma^2 )^{-3/2} \{ \exp( \frac{- \varepsilon + \Phi_0}{\sigma^2} ) - 1 \},\label{eq:king}
 \end{eqnarray}
where, $\rho_1$, $\sigma$, $\Phi_0$ are fitting parameters.
In figure 3, figure 4 and figure 5, distribution functions of the 
best fit King-models are showed by the short dash lines.

Then we define $D^2$ as follows and use it for the estimation of 
the deviation from Lynden-Bell distribution.
 \begin{eqnarray}
 D^2 \equiv  \Bigr( \sum_{i=1}^{N_b} n_i \Bigl)^{-1}  \sum_{i=1}^{N_b} n_i \Biggr| \frac{\bar{f}_{data}(\varepsilon_i) - \bar{f}_{fit}(\varepsilon_i) } { \bar{f}_{fit}(\varepsilon_i) } \Biggl|^2,   \label{eq:d2}
 \end{eqnarray}
where, $\bar{f}_{data}$ is the distribution function from simulation data, 
$\bar{f}_{fit}$ is the best fit function of the data 
in the form of (\ref{eq:king}), $N_b$ is the number of bins (in this paper
$N_b$ = 100) and  $n_i$ is the number of shells contained in the $i$-th bin.
Fluctuations from the finiteness of shell number are proportional to 
square root of shell number. Therefore, the summation is carried out 
weighted with $n_i$ in the numerator of equation (\ref{eq:d2}).

In table 3, fitting parameters ( $\rho1,\sigma^2,\Phi_0$ ),  $c$,
which is the concentration parameter of the corresponding King model 
obtained from the fitting result and $D^2$ of each model are listed. 
Compared between Model A-0 and B-0, both of which are without external force,
$D^2$ of B-0 is smaller than that of A-0. Initial condition of model B-0 is
farther from dynamical equilibrium state than that of A-0, so time variation of
gravitational potential is larger and relaxation is more effective.
In all models whose initial virial ratios are 0.5 with external force 
(from A-1 to A-9), $D^2$ becomes smaller than that of model A-0 which has 
the same initial condition but without external force. The similar tendency 
can be seen in almost models whose initial virial ratio is 0.1 
(from B-0 to B-1) except B-4 and B-7.

\section{Discussion}

We examined the violent relaxation process of spherical stellar systems
and the influence on it of time variation of external gravitational potential
fields by using shell models. It is found that time variation of the mean 
potential field mainly caused by external gravitational field makes
distribution function closer to Maxwellian. Especially in some models
(A1, A4, A7, B8, B9) the deviations from the King model become quite small.
Though it is not clear why the deviations are small in these models, 
it shows that the violent relaxation is quite efficient under some conditions.

So, if time variation of mean gravitational potential fields continues for
a long enough time, systems can reach Lynden-Bell distribution.

It is rare that no gravitational influence of other galaxies is
exercised on galaxies. Especially in galaxies in clusters, this effect
plays an important role on their dynamical evolution. Though it is not clear
the real gravitational perturbations are how much effective to cause 
the violent relaxation, there are possibilities that 
the distribution function of the real elliptical galaxies become close to
the King models by the external forces.

\bigskip
\bigskip

We would like to thank T.Hayashi for his contribution to developing numerical
code and M. Shimada for helpful discussions. 
Numerical computations in this work were carried out with workstations
at Yukawa Institute for Theoretical Physics.
This work is in part supported by Research Fellowships of the Japan Society
for the Promotion of Science for Young Scientists (M.T.).

\setcounter{section}{0}
\renewcommand{\thesection}{Appendix \arabic{section}.}		
\renewcommand{\theequation}{A\arabic{section}.\arabic{equation}}

\section{ Estimation of Two-Body Relaxation Time of Shell Systems}
\setcounter{equation}{0}

We assume that there are $N$ shells which have the same mass of $m$ .
Then the mean energy of each shell $\bar{ \varepsilon }$ is as follows from 
the Virial theorem.
 \begin{eqnarray}
 \bar{ \varepsilon } \sim - \frac{ G N m^2 }{ \bar{R} },
 \end{eqnarray}
where $G$ is the gravitational constant and $\bar{R}$ is a typical radius
of the system.
And the mean energy change of each shell in one shell-crossing-event is 
 \begin{eqnarray}
 \vert \bar{\Delta \varepsilon} \vert \sim \frac{G m^2}{ \bar{R} }.
 \end{eqnarray}

Relaxation process can be regarded as one-dimensional random walks in the 
energy space whose one step-size is equal to 
$\vert \bar{\Delta \varepsilon} \vert$ .
Since the system can be regarded as relaxed when the variance of 
displacement in the energy space becomes comparable to the square of 
typical energy of each shell,
number of crossing-event required for relaxation, $n_{rel}$, is calculated
as follows.
 \begin{eqnarray}
  \bar{ \varepsilon }^2 \sim n_{rel} { \vert \bar{\Delta \varepsilon} \vert }^2, \\ n_{rel} \sim ( \frac{\bar{\varepsilon}}{\bar{\Delta \varepsilon} } )^2 \sim N^2.
 \end{eqnarray}

Then relaxation time of shell systems, $t_{rel}$, is 
 \begin{eqnarray}
 t_{rel} \sim n_{rel} \frac{\bar{r}}{\bar{v}} \sim N \frac{\bar{R}}{\bar{v}} \sim N t_{cr},
 \end{eqnarray}
where $\bar{r} = \bar{R} / N$ is the mean separation between shells and 
$\bar{v}$ is the mean velocity of shells.

\section{Changes of Length and Time Scale through Collapse}
\setcounter{equation}{0}

  We assume that the initial total mass is $M$, the initial typical length 
scale is $R_0$, and that the initial velocity dispersion is $v_0^2$. 
Then total kinetic energy, $T_0$, total potential energy, $W_0$, and 
virial ratio, $V_0$, of initial states are,
 \begin{eqnarray}
  T_0 &=& \frac{1}{2} M v_0^2, \\
  W_0 &=& - \frac{ G M^2 }{ R_0 }, \\
  V_0 &=& \biggl | \frac{ 2 T_0}{W_0} \biggr | = \frac{v_0^2 R_0}{G M},
 \end{eqnarray}
where $G$ is the gravitational constant.

 When the system reached virial equilibrium after the collapse, we let the 
typical length scale $R$, velocity dispersion $v^2$, and the total mass 
$M(1-x)$, where $x$ is the escape rate of mass. Then total kinetic energy, 
$T$, total potential energy, $W$, and virial ratio, $V$, of the 
bounded portion of the system are,
 \begin{eqnarray}
  T &=& \frac{1}{2} M (1-x) v^2, \\
  W &=& - \frac{GM^2}{R} (1-x)^2, \\
  V &=& 1 = \frac{v^2 R}{G M (1-x)}.
 \end{eqnarray}

If we let the energy which escaped mass possess $\Delta E$, according to energy
conservation law,
 \begin{eqnarray}
  T_0 + W_0 = T + W + \Delta E,
 \end{eqnarray}
then,
  \begin{eqnarray}
   v^2 &=& \frac{G M}{R_0} (2-Vr_0) (1-x)^{-1} (1+\frac{\Delta E}{|E_0|}) \\
       &=& v_0^2 ( -1 + \frac{2}{Vr_0}) (1-x)^{-1} (1+\frac{\Delta E}{|E_0|}), \\
   R &=& R_0 (2-Vr_0)^{-1} (1-x)^2 (1+\frac{\Delta E}{|E_0|})^{-1},
\label{eq:r} \\
   t &=& \frac{1}{\sqrt{GM}} ( \frac{R_0}{2-Vr_0} )^{2/3} (1-x)^{5/2} (1+\frac{\Delta E}{|E_0|})^{-3/2} \label{eq:t} \\
     &=& t_0 \frac{Vr_0^{1/2}}{(2-Vr_0)^{3/2}} (1-x)^{5/2} (1+\frac{\Delta E}{|E_0|})^{-3/2},
  \end{eqnarray}
 where $|E_0| = G M^2 (1-Vr_0/2)/R_0$ is the absolute value of initial total
energy of the system, $t = R/v$  is the typical timescale of the system after 
the system has collapsed, and $t_0 = R_0/v_0$ is that of initial state.
 
 In our simulation, when $V_0=0.5$, $x$ and $\Delta E / |E_0|$ is much smaller
than unity and can be negligible. On the other hand, when $V_0 = 0.1$,
$x=0.32$ and $\Delta E / | E_0 | = 0.45$ . With our unit ( $G = M = 1$ )
and the initial conditions described in section 3, when $V_0 = 0.5$, $R = 1.34$
 and $t=1.53$ from (\ref{eq:r}) and (\ref{eq:t}). When $V_0 = 0.1$, $R=0.494$ and $t=0.345$.

\newpage
{\bf Reference}

Cuperman S., Goldstain S., Lecar M. 1969, MNRAS 146, 161

Funato Y., Makino J., Ebisuzaki T. 1992a, PASJ  44, 291

Funato Y., Makino J., Ebisuzaki T. 1992b, PASJ  44, 613

H\'enon M. 1964, Ann. Astrophys. 27, 83

Goldstein S., Cuperman S., and Lecar M. 1969, MNRAS 143, 209

Hohl F., Campbell D. W. 1968, AJ 73, 611 

Lynden-Bell D. 1967, MNRAS 136, 101

Yamashiro T., Gouda N., Sakagami M. 1992, Prog. Theor. Phys. 88, 269

\newpage

\begin{center}
 Table 1. Parameters of Models \\
\bigskip
 \begin{tabular}{ccccc}
  \hline\hline
  Model & Initial V.R. & External P.F. &  $P / \bar{t}$ & $\lambda / \bar{R}$ \\
  \hline
  A-0   &   0.5        &      No       &     --         &  --       \\
  A-1   &   0.5        &      Yes      &     0.5        &  0.5      \\
  A-2   &   0.5        &      Yes      &     0.5        &  1.0      \\
  A-3   &   0.5        &      Yes      &     0.5        &  2.0      \\
  A-4   &   0.5        &      Yes      &     1.0        &  0.5      \\
  A-5   &   0.5        &      Yes      &     1.0        &  1.0      \\
  A-6   &   0.5        &      Yes      &     1.0        &  2.0      \\
  A-7   &   0.5        &      Yes      &     2.0        &  0.5      \\
  A-8   &   0.5        &      Yes      &     2.0        &  1.0      \\
  A-9   &   0.5        &      Yes      &     2.0        &  2.0      \\
  B-0   &   0.1        &      No       &     --         &  --       \\
  B-1   &   0.1        &      Yes      &     0.5        &  0.5      \\
  B-2   &   0.1        &      Yes      &     0.5        &  1.0      \\
  B-3   &   0.1        &      Yes      &     0.5        &  2.0      \\
  B-4   &   0.1        &      Yes      &     1.0        &  0.5      \\
  B-5   &   0.1        &      Yes      &     1.0        &  1.0      \\
  B-6   &   0.1        &      Yes      &     1.0        &  2.0      \\
  B-7   &   0.1        &      Yes      &     2.0        &  0.5      \\
  B-8   &   0.1        &      Yes      &     2.0        &  1.0      \\
  B-9   &   0.1        &      Yes      &     2.0        &  2.0      \\
  \hline
 \end{tabular}
\end{center}  

\newpage

 Table 2. Ratio of mass whose energy is more than zero in stationary states.\\
\bigskip
\begin{center}
 \begin{tabular}{cccc}
  \hline\hline
  Model &    Rate      &   Model & Rate   \\
  \hline
  A-0   &   0.0003     &  B-0    & 0.33   \\
  A-1   &   0.78       &  B-1    & 0.85   \\
  A-2   &   0.11       &  B-2    & 0.39   \\
  A-3   &   0.0029     &  B-3    & 0.31   \\
  A-4   &   0.41       &  B-4    & 0.60   \\
  A-5   &   0.78       &  B-5    & 0.82   \\
  A-6   &   0.20       &  B-6    & 0.43   \\
  A-7   &   0.18       &  B-7    & 0.42   \\
  A-8   &   0.53       &  B-8    & 0.64   \\
  A-9   &   0.75       &  B-9    & 0.80   \\
  \hline
 \end{tabular}
\end{center}

\newpage

 Table 3. Fitting parameters ( $\rho_1$, $\sigma^2$, $\Phi_0$ ) and concentration parameter, $c$, and $D^2$ of each model  \\
\bigskip
\begin{center}
 \begin{tabular}{cccccc}
  \hline\hline
  Model & $\rho_1$ & $\sigma^2$ & $\Phi_0$ & $c$ &  $D^2$  \\
  \hline
  A-0   & $2.43 \times 10^{-4}$ &   0.181   &  $-2.25 \times 10^{-2}$ & 1.52 & 1.26   \\
  A-1   & $6.31 \times 10^{-9}$ &  $9.80 \times 10^{-3}$  & $-1.56 \times 10^{-3}$ & 2.44 & 0.360   \\
  A-2   & $6.05 \times 10^{-5}$ &   0.102   &  $-1.29 \times 10^{-2}$ & 1.54 &  0.725    \\
  A-3   & $2.13 \times 10^{-4}$ &   0.168   &  $-2.10 \times 10^{-2}$  & 1.51 &   0.954 \\
  A-4   & $4.32 \times 10^{-7}$ &   $2.71 \times 10^{-2}$ & $-4.18 \times 10^{-3}$ & 2.05 & 0.427  \\ 
  A-5   & $5.71 \times 10^{-9}$ & $9.80 \times 10^{-3}$ & $-1.67 \times 10^{-3}$ & 2.33 &  0.588 \\
  A-6   & $2.57 \times 10^{-6}$ & $4.27 \times 10^{-2}$ & $-6.15 \times 10^{-3}$ & 1.86 &  1.19 \\   
  A-7   & $1.60 \times 10^{-5}$ & $7.40 \times 10^{-2}$ & $-8.95 \times 10^{-3}$ &1.67  &  0.363 \\
  A-8   & $4.81 \times 10^{-6}$ & $5.32 \times 10^{-2}$ & $-6.62 \times 10^{-3}$ & 1.61 &  0.506 \\ 
  A-9   & $2.22 \times 10^{-8}$ & $1.33 \times 10^{-2}$ & $-2.11 \times 10^{-3}$ & 2.13 &  0.852 \\
  B-0   & $1.80 \times 10^{-2}$ & 0.541 & $-6.48 \times 10^{-2}$ & 1.46 &  0.923       \\
  B-1   & $1.20 \times 10^{-9}$ & $7.19 \times 10^{-3}$ & $-1.52 \times 10^{-3}$ & 3.01 & 0.633  \\
  B-2   & $2.12 \times 10^{-3}$ & 0.227 & $-2.67 \times 10^{-2}$ & 1.45 & 0.720        \\
  B-3   & $1.50 \times 10^{-2}$ & 0.485 & $-5.60 \times 10^{-2}$ & 1.41 &  0.726  \\
  B-4   & $1.55 \times 10^{-5}$ & $5.32 \times 10^{-2} $ & $-8.65 \times 10^{-3}$ & 2.11 & 1.04  \\
  B-5   & $9.33 \times 10^{-9}$ & $1.04 \times 10^{-2}$ & $-2.31 \times 10^{-3}$ & 3.01 & 0.534 \\
  B-6   & $6.17 \times 10^{-4}$ & 0.149 & $-1.82 \times 10^{-2}$ & 1.49 & 0.779  \\
  B-7   & $4.93 \times 10^{-4}$ & 0.151 & $-1.81 \times 10^{-2}$ & 1.67 & 1.29      \\
  B-8   & $9.48 \times 10^{-7}$ & $3.32 \times 10^{-2}$ & $-6.35 \times 10^{-3}$ & 2.60 & 0.387  \\
  B-9   & $2.12 \times 10^{-8}$ & $1.21 \times 10^{-2}$ & $-2.48 \times 10^{-3}$ & 2.78 & 0.250  \\
  \hline
 \end{tabular}
\end{center}

\newpage
\begin{center}
\bf{Figure captions}
\end{center}

\begin{description}
\item[Fig. 1.] Time variation of virial ratio of model A-0. The system 
approached stationary state within a few crossing times and variation 
of virial ratio was damped.

\item[Fig. 2.] Same as figure 1, but for model B-0. the mean of virial ratio 
of stationary state is greater than $1.0$ (about $1.35$) because the 
existence of shells which have energy of more than zero cannot be negligible.

\item[Fig. 3.] Distribution function of stationary states of model A-0
(by the solid line) and that of the best fit King-model 
(by the short dash line).

\item[Fig. 4.] Same as figure 3, but for model B-0.

\item[Fig. 5.] Same as figure 3, but for model A-1.
$\bar{f} ( \varepsilon )$ obeys Maxwellian. The deviation from Maxwellian 
of lower energy region is due to the statistical fluctuation because
number of shells in this energy range is rather small.
\end{description}

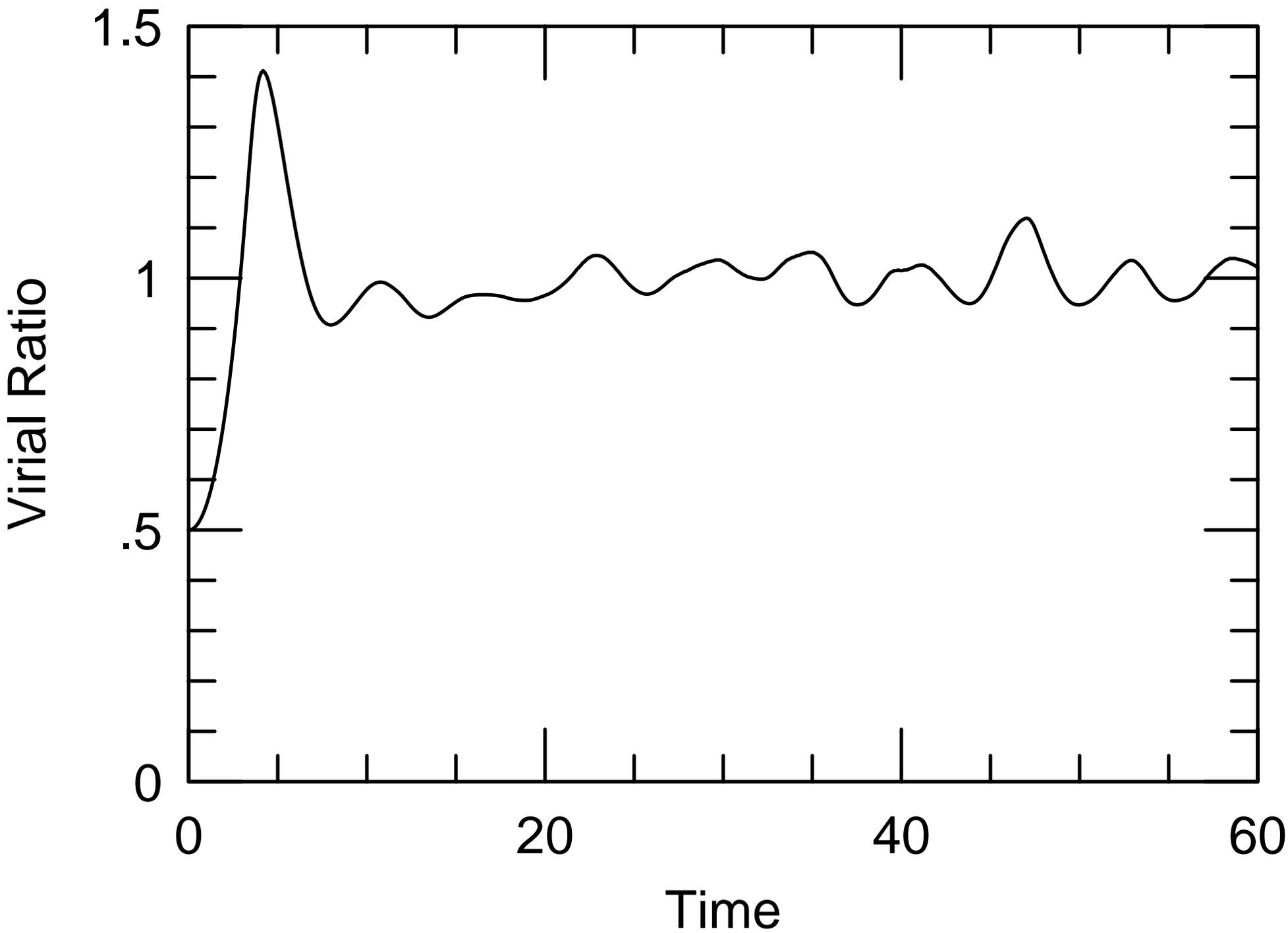
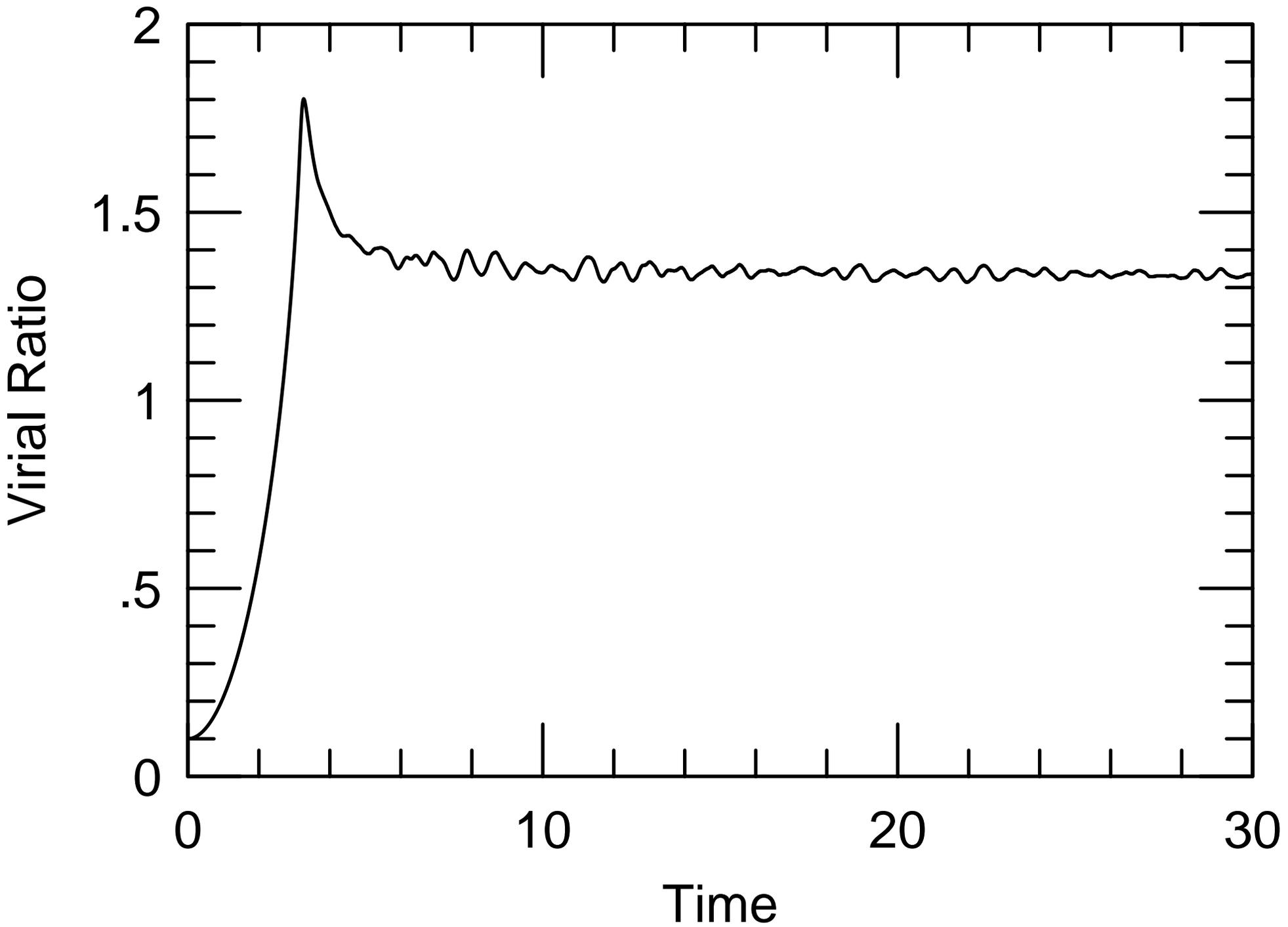
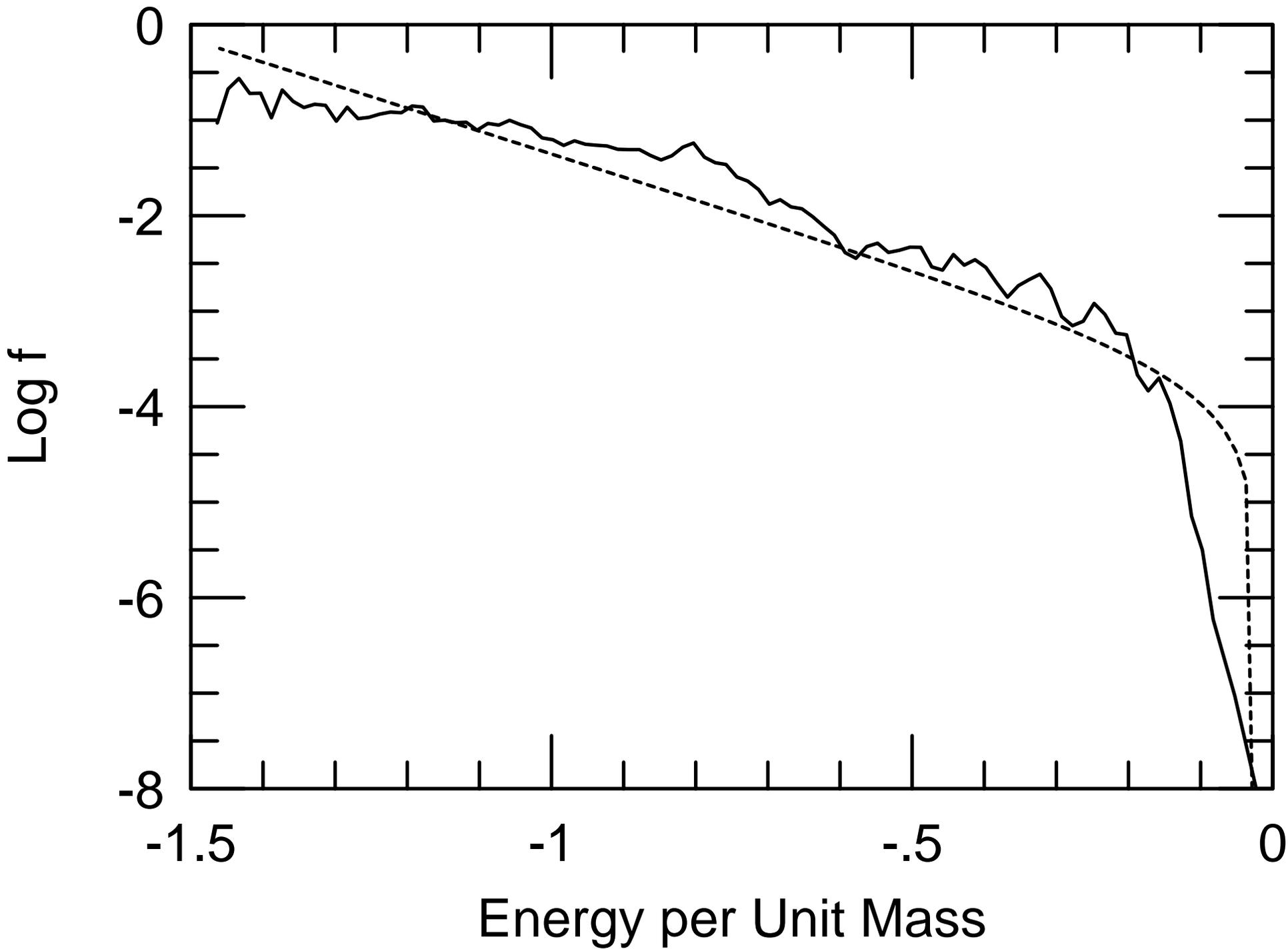
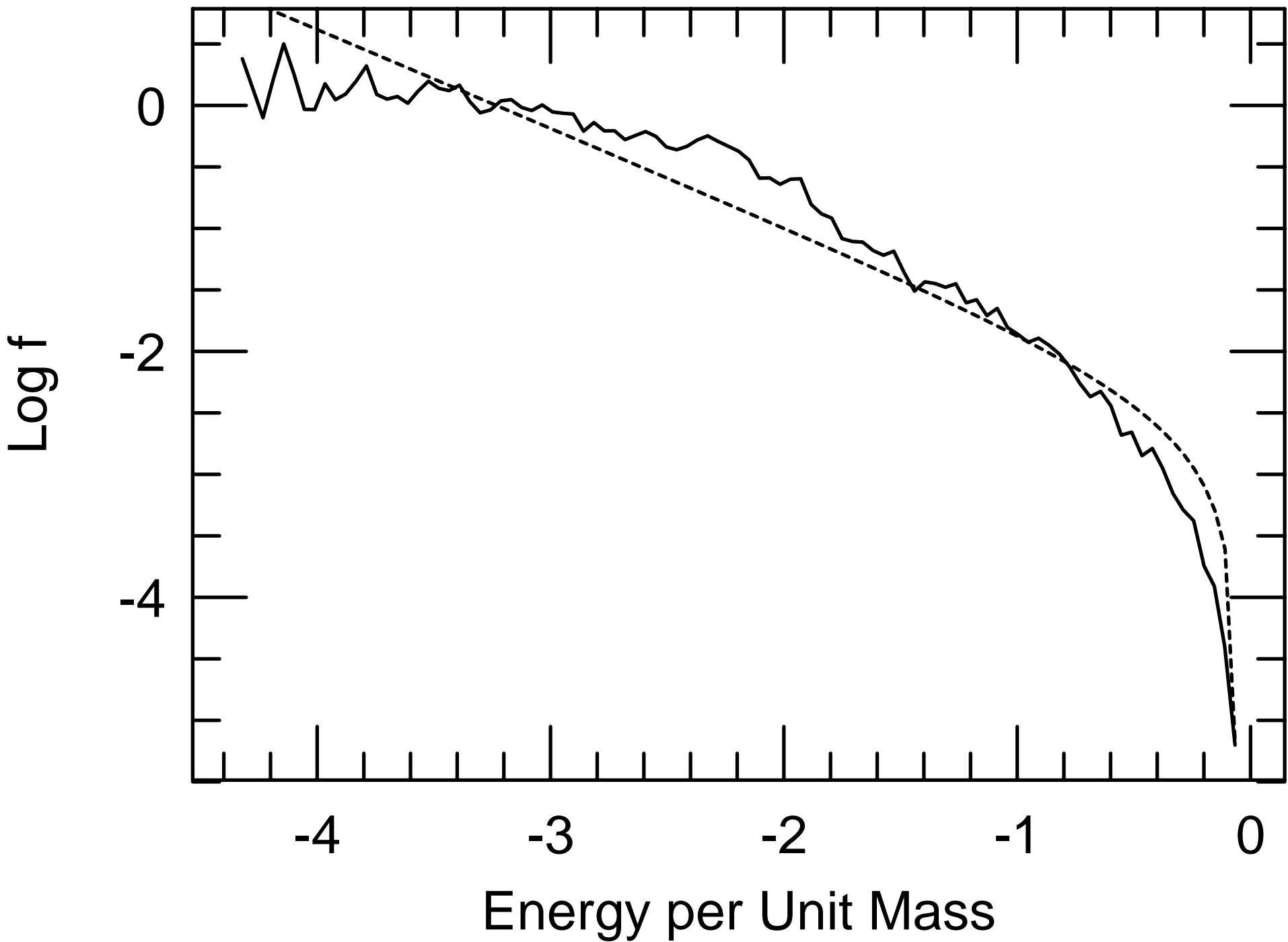
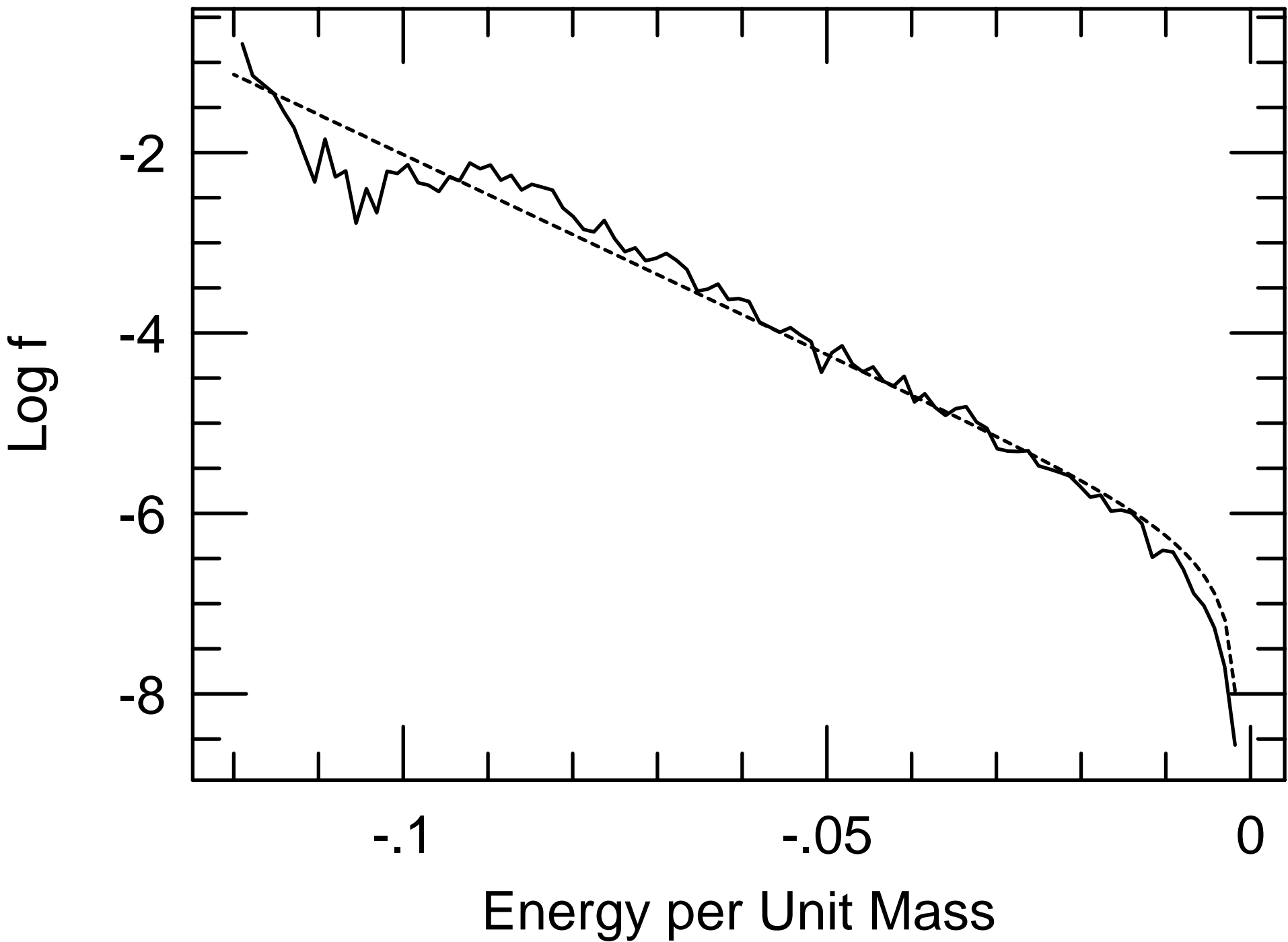
\end{document}